\documentclass[preprint,12pt]{elsarticle}

\usepackage{amssymb}
\usepackage{lineno}
\usepackage{subfigure}
\usepackage[normalem]{ulem}
\usepackage{subfigure}
\usepackage[cmex10]{amsmath}
\usepackage{algorithmic}
\usepackage{array,tabularx,tabulary,booktabs} 
\usepackage{multirow}
\usepackage{amsfonts,amssymb,amsthm,mathtools}
\usepackage{icomma} 
\usepackage{xcolor}

\usepackage{calc, graphicx}  
\usepackage{epstopdf}  

\usepackage{hyperref}
\usepackage{fancyhdr}

\fancypagestyle{firststyle}
{
   \fancyhf{}
   
   \fancyhead[C]{\tiny{This is the post-print of the following article: O. Vikhrova, S. Pizzi, A. Iera, A. Molinaro, K. Samuylov and G. Araniti, ``Group-based Delivery of Critical Traffic in Cellular IoT Networks," in Elsevier Computer Networks, doi: 10.1016/j.comnet.2020.107563.\\Article has been published in final form at: \url{https://www.sciencedirect.com/science/article/pii/S1389128620312111}. \\\copyright~2020 Elsevier B.V. All rights reserved.}}
}

\journal{Computer Networks}

\begin{document}

\begin{frontmatter}



\title{Group-based Delivery of Critical Traffic \\in Cellular IoT Networks}

 \author[1]{Olga Vikhrova\corref{cor1}}
	\ead{olga.vikhrova@unirc.it}
	\author[1]{Sara Pizzi} \ead{sara.pizzi@unirc.it}
	\author[1]{Antonella Molinaro} \ead{antonella.molinaro@unirc.it}
	\author[2]{Antonio Iera} \ead{antonio.iera@dimes.unical.it}
	\author[3]{Konstantin Samuylov} \ead{samuylov-ke@rudn.ru}
	\author[1]{Giuseppe Araniti} \ead{araniti@unirc.it}
	
\cortext[cor1]{Corresponding author}
\address[1]{DIIES Dept., University Mediterranea of Reggio Calabria, Italy}
\address[2]{DIMES Dept., University of Calabria, Italy}
\address[3]{Peoples' Friendship University of Russia (RUDN University), Russia}

\begin{abstract}
Fifth generation (5G) networks are expected to connect a huge number of Internet of Things (IoT) devices in many usage scenarios. The challenges of typical massive IoT applications with sporadic and short packet uplink transmissions are well studied, while not enough attention is given to the delivery of content of common interest, such as software/firmware updates and remote control, towards IoT devices in emerging point-to-multipoint scenarios. 
Moreover, the delivery of delay-sensitive IoT traffic is not sufficiently addressed in the literature. In this work we (i) identify the drawbacks of the current Single-Cell Point-to-Multipoint (SC-PTM) solution for unplanned critical traffic delivery in cellular IoT (cIoT) networks, and (ii) propose paging and multicast schemes for a fast distribution of critical updates after, e.g., bug fixes or system failures. We benchmark the performance of the proposed paging scheme against similar solutions available in the literature. Our extended SC-PTM framework is energy efficient and guarantees low service latency, as demonstrated both analytically and by simulations.
\end{abstract}


\end{frontmatter}

\thispagestyle{firststyle}


\section{Introduction}
\label{sec:1}

Fifth-generation (5G) networks are expected to connect a huge number of heterogeneous devices. Differently from previous generations of cellular networks, 5G strongly focuses on massive Machine-Type Communications (MTC) and Internet of Things (IoT), addressing both massive MTC (mMTC) and Ultra-Reliable and Low-Latency Communication (URLLC) use cases~\cite{ITU-R}. 
Many of the emerging IoT use cases move the focus from sporadic data transmissions in the uplink (UL) direction -- such as smart gas-metering devices that wake up once a day to send the consumption reports to the gas-metering network -- to simultaneous data delivery from network to multiple receivers in the downlink (DL). The latter case includes software/firmware updates, system configuration changes, and remote device control~\cite{3GPP26850}. 

Point-to-Multipoint (PTM) communication is the key technology in such scenarios, because of its capability to feed a theoretically unlimited number of devices in a single transmission~\cite{5G_araniti,Rinaldi}. 
The $3^{\text{rd}}$ Generation Partnership Project (3GPP) specified the subscription-based Multimedia Broadcast Multicast Service (MBMS) architecture
to provide a way for the network to deliver the content of interest towards multiple receivers over a large number of cells~\cite{3GPP26346}.
Successively, the Single-Cell Point-to-Multipoint (SC-PTM) operation mode was introduced in Release 13 to support multicast data delivery in a single cell. In Release 14, it was enabled for Narrowband IoT (NB-IoT) and Long Term Evolution for Machines (LTE-M), which are recognized as 5G solutions that meet technical requirements of large-scale mMTC scenarios~\cite{ITU-R} and ensure coexistence with the 5G New Radio (NR)~\cite{gsma}. 

In conventional multicast scenarios, devices create a \textit{multicast group} by subscribing to the content of interest and wait for the service announcement when the content is available for download. The service announcement stage usually runs for a long time to ensure that all devices in the group get ready for the content reception when multicast transmission starts.

In this paper, we focus on the challenging use case of a critical update dissemination towards a large number of IoT devices as a consequence of  critical bug fixes or system reconfiguration because of a failure.
Since devices are not aware of message arrival, network needs to send a paging message first to notify them of incoming data. The multicast group can not be created in advance and multicast transmissions can not be scheduled as in the example above because the critical content must be delivered to IoT devices as soon as possible.

\subsection{Related work and contribution of this paper}

The need for a customer-driven group formation for PTM services in cellular IoT (cIoT) has been early discussed in~\cite{5G_araniti}. MTC devices usually operate in a limited regime to save battery, they sporadically wake up to perform routine tasks and upload only few bytes to an application server. The eMBMS is a Human-Type Communication (HTC) oriented technology that assumes all end-points being under human control, which is not the case of MTC. 

The trade-off between device availability for network-originated data and device energy consumption is well covered in the literature. For instance, the work in ~\cite{DRX2} discusses the impact of device active and sleep periods on the expected battery life cycle. In~\cite{Oh2017}, device energy consumption under different active and sleep intervals and when varying traffic rates is analyzed by assuming unicast DL transmissions. The results demonstrate that both very short and very long intervals between paging indication and DL traffic arrival lead to an increase in device energy consumption. Similar results have been reported in~\cite{Sultania2018} for more types of traffic and use cases. Device grouping is exploited in~\cite{Xu2018} to improve the energy consumption of IoT devices with similar UL traffic pattern and Quality of Service (QoS). The grouping algorithm helps to avoid congestion in the UL when a huge number of devices try to access the network after receiving paging indication. 
However, mentioned works are mainly focused on the issue of \textit{paging}, either to improve long-term device energy consumption with regular traffic or, alternatively, to reduce device collision rate in the UL. In our proposal, we address both paging and multicast traffic delivery aspects. 

In a previous work~\cite{PIMRC}, we proposed three different strategies to group IoT devices for the reception of multicast traffic. The first strategy is meant to group all relevant devices into a single group and schedules SC-PTM transmission when the last device of the group enters the Radio Resource Control (RRC) connected state joining the multicast group. According to the second strategy, devices are split into multicast groups of equal size; connected devices wait for the SC-PTM transmission until the group is formed. In the last strategy, we proposed to schedule identical multicast transmissions any moment when devices are ready for the data reception, i.e. any number of devices may fall into the multicast group. We considered only legacy paging strategy, defined by 3GPP, to notify devices of the multicast service; according to it, not more than 16 devices can be reached by one paging transmission~\cite{SP}.

In~\cite{BMSB}, we discussed the necessary improvements of the SC-PTM service announcement and proposed a new grouping solution for the multicast reception of critical content, considering the drawbacks of the strategies from~\cite{PIMRC}. In the new strategy, the network schedules SC-PTM transmissions in a fixed interval named \textit{critical interval}. However, we did not discuss how this interval should be adjusted. We extended the analysis with two enhanced paging strategies from the reference literature, namely \textit{Group paging} (GP)~\cite{GP}, which allows addressing any number of devices in one paging message, and \textit{enhanced Group paging} (eGP)~\cite{Condoluci} where paging is sent out over fixed intervals to a group of devices.

Solution for paging in~\cite{HGP} improves device's battery life cycle at the expense of a very long service delay that is unacceptable for critical applications. Authors in~\cite{EE_GP} obtained the optimal size for a paging group based on the limited capacity of the \textit{Random Access} (RA) followed by paging. However, none of the mentioned works, except for~\cite{Condoluci}, takes into account the impact of paging on multicast efficiency. For this reason, we propose a new paging solution that leaves from the general idea of the paging approaches proposed in~\cite{Condoluci} and~\cite{EE_GP}, but reinforces our SC-PTM transmission scheme for the delay critical IoT applications.

Before us, authors in~\cite{Feltrin} analysed the performance of the firmware updates over unicast and PTM links for NB-IoT. The work~\cite{Tsoukaneri} deals with the resource allocation problem for the multicast transmission in the presence of unicast traffic. Both works lack an analytic approach and solutions for paging, which are contributions of our work.
Our paging and device grouping solutions have been evaluated analytically and validated by extensive simulations.

The main contributions of this work are: 
\begin{itemize}
	\item[--] A multicast framework for critical cIoT services that helps to avoid long legacy service announcement procedure, efficiently pages devices and schedules SC-PTM transmissions.
	\item[--] A new paging strategy that properly adjusts the paging interval and size of the paging groups to improve the probability of content reception and reduce delay of SC-PTM services.
	\item[--] An analytical framework that accurately models all the phases involved in SC-PTM service provision, such as paging, system configuration for the SC-PTM reception and multicast transmission itself.
	\item[--] An extensive numerical analysis with device and network oriented metrics and different payloads of the multicast traffic that may represent very short commands, alerts and small bug fixes.
\end{itemize}
We also discuss minor but necessary changes in some messages of the RA stage, not addressed in~\cite{BMSB}.

The rest of the paper is organized as follows. 
In section \ref{sec:back}, we give the background on paging and RA procedures and explain the necessary changes for SC-PTM to make delivery of critical traffic in cIoT feasible.
The details of our proposal are given in section~\ref{sec:model}, while numerical results are discussed in section~\ref{sec:results}. Conclusive remarks are given in the last section.

\section{Setting the scene}
\label{sec:back}

\subsection{Paging and Random access procedures}
\label{sec:PRA}

The individual activity pattern of cellular IoT devices is determined by their duty cycle, alternating short \textit{connected} and long \textit{idle} periods. Therefore, \textit{paging} is needed to notify the arrival of DL data when device is in idle mode. The duration of active and idle intervals is defined by the discontinuous reception (DRX) strategy. In 3GPP Release 13, an extended DRX (eDRX) strategy has been introduced, which, compared to the Power Saving Mode (PSM), allows IoT devices to remain idle for longer period, save energy, and improve their response time in applications with network-originated traffic.

After an inactivity period since the last transmission, defined by the \textit{Inactivity Timer}, the device turns the receiver circuitry off and only periodically listens to the Paging Radio Network Temporary Identifier (P-RNTI) indication in the Physical Downlink Control Channel (PDCCH). Note that LTE-M and NB-IoT use ad-hoc designed MTC PDCCH (MPDCCH) and Narrowband PDCCH (NPDCCH)~\cite{3GPP45820}. For the sake of brevity, we omit to specify the exact name of the different physical LTE-M and NB-IoT channels. 
It wakes up for the \textit{onDuration} time to receive the paging message and to look up for its identifier (ID) in the \textit{paging records} list. If the device finds the appropriate record then it follows the instruction from the paging message, otherwise it turns back to sleep~\cite{3GPP36304}.

Two parameters help to define when a device is available for a PTM service: the Paging Frame (PF) and the Paging Opportunity (PO) indicating the radio frame and subframe when the device must listen to the paging indication in the PDCCH. For NB-IoT devices, the concept of Paging Narrowband (PNB) replaces PO to indicate not only the subframe but also the narrowband where paging indication can be received. For simplicity, we refer to PO only, including also PNB in this term. 
Network can address several devices at a time if they listen to the same PO at the same PF including their IDs into the paging record list. However, the number of paging records in one message is limited~\cite{SP}. Alternatively, it may address devices by their Group ID (GID)~\cite{GP} if assigned previously.

\begin{figure}[tbh]
	\centering
	\includegraphics[scale=0.6]{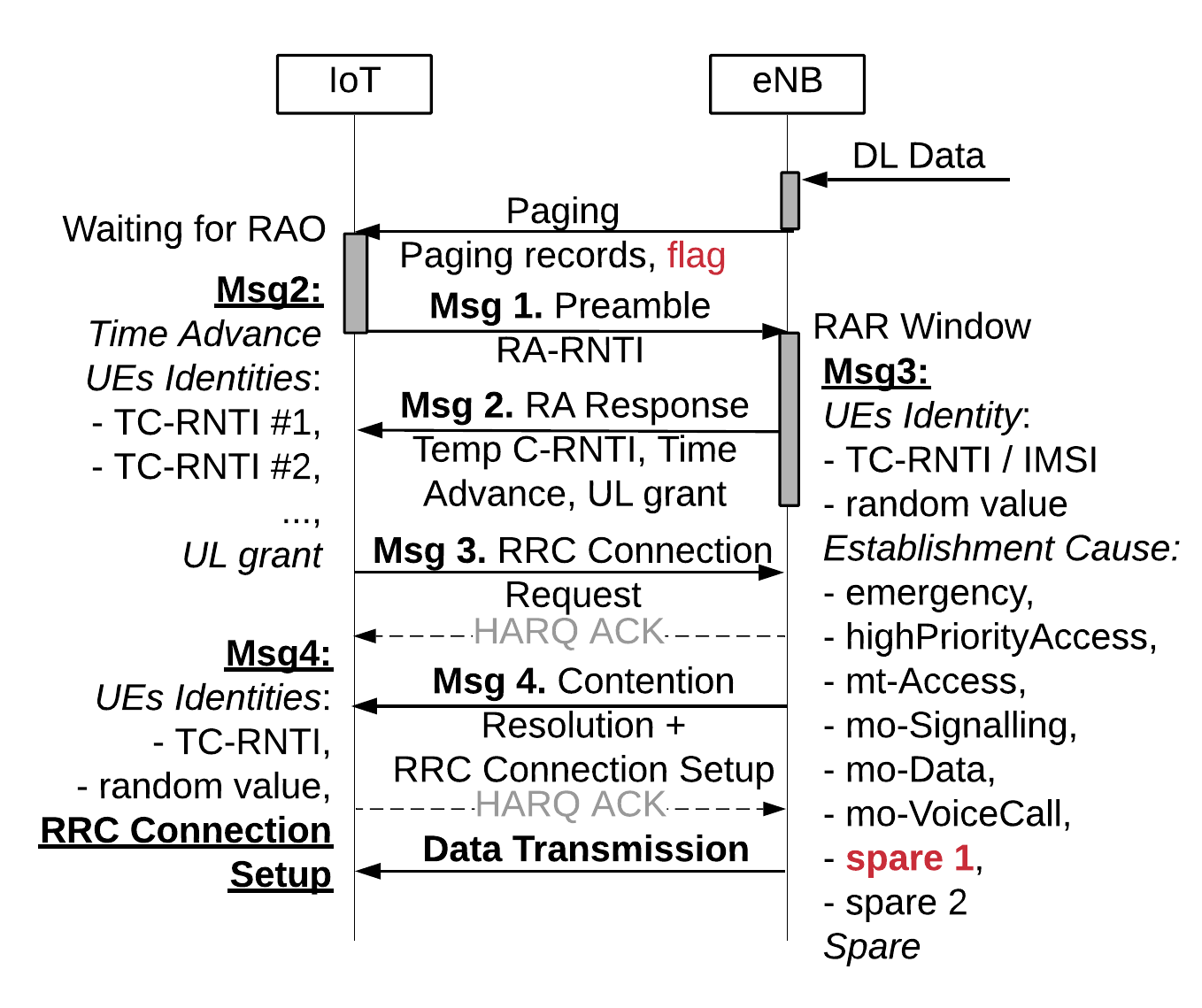}
	\caption{3GPP complaint RA procedure.}
	\label{fig:RACH1}
\end{figure}

Once devices are awake, they are immediately ready for the data reception, but they are unable to transmit. In order to send a feedback to the network, devices must synchronize with the BS and request resource allocation for the subsequent transmission in uplink. Without feedback from devices, the BS has to continuously broadcast data at the lowest rate over a long period of time to ensure that all devices get the content.

To synchronize with the BS, a device initiates the RA procedure, as illustrated in Fig.~\ref{fig:RACH1}, by sending a randomly chosen preamble (\textit{Msg1}) over the physical random access channel (PRACH) scheduled at specific random access opportunities (RAOs), defined by the PRACH configuration index. 
If the BS successfully decodes Msg1, then it replies with the RA response (RAR) message (\textit{Msg2}), including the Temporary Cell-Radio Network Temporary Identifier (TC-RNTI), the timing advance information for synchronization purpose, and a UL grant for the next message transmission in the physical uplink shared channel (PUSCH). Then the device sends the Radio Resource Control (RRC) connection request (\textit{Msg3}) and specify the \textit{Establishment Cause}.
If the BS decodes Msg3 it replies with Contention Resolution message (\textit{Msg4}) using identifiers from the Msg3. If both TC-RNTI and UE Identity equal to the TC-RNTI and UE Identity that the device included in Msg3, the RA stage is successfully completed.

Preamble retransmissions can be triggered due to the lack of resources for Msg2 transmission or due to collisions upon  Msg3 transmission. These retransmissions are the events that contribute most to the access delay and may cause device access failure. After a failed RA attempt, the device waits for a backoff interval and then retries with a preamble transmission. When the maximum number of retransmissions is reached, the device is considered to be unable to connect to the network due to poor link conditions, and may go back to the idle mode. 

\subsection{Multicast Framework for critical IoT applications}
\label{sec:SCPTM}

\begin{figure}[htbp]
	\centering
	\includegraphics[scale=0.3]{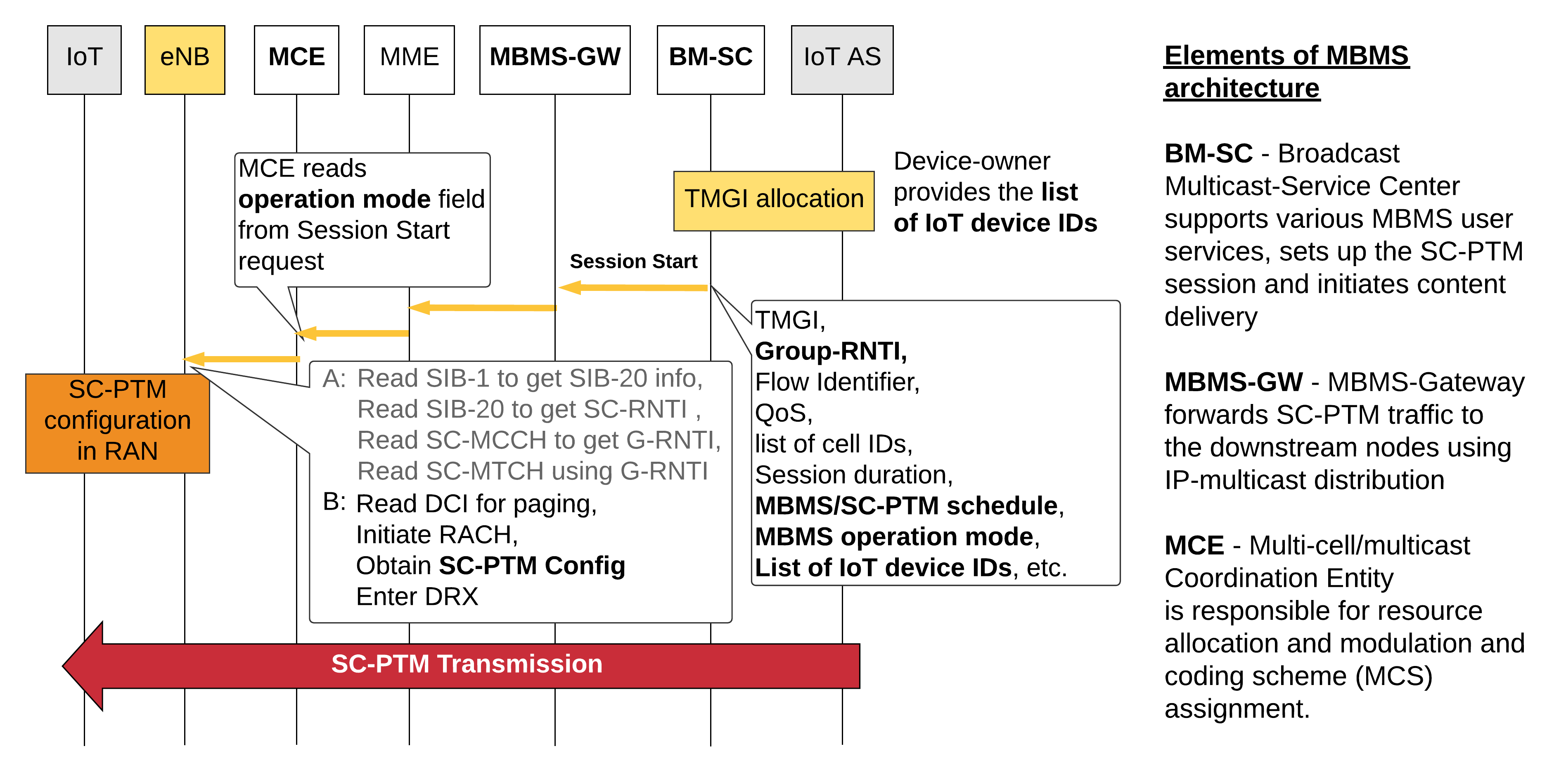}
	\caption{Standard (Option A) and proposed (Option B) scheme to deliver SC-PTM traffic towards cIoT devices.}
	\label{fig:MtMSpro}
\end{figure}

SC-PTM reuses the MBMS architecture but utilizes supplementary radio bearer service. 
SC-PTM control and data are transferred in the dedicated Single-Cell Multicast Control Channel (SC-MCCH) and Single-Cell Multicast Traffic Channel (SC-MTCH) respectively. These two channels dynamically are mapped to the Physical Downlink Shared Channel (PDSCH) with prior indication in the PDCCH~\cite{3GPP26346},~\cite{3GPP36331}. Each multicast session has a unique Temporary Mobile Group Identity (TMGI) in core and radio access segments. Similar to paging, SC-PTM control and traffic transmissions are indicated by SC-PTM RNTI (SC-RNTI) and Group-RNTI (G-RNTI) in DCI respectively. Once a device gets TMGI, G-RNTI and scheduling information for the SC-PTM transmission (i.e., scheduling period, scheduling window and start offset), it can receive the content, as shown in the Fig.~\ref{fig:MtMSpro}.

3GPP-based SC-PTM for cIoT is only supported in idle mode. To this end, a new System Information Block Type 20 (SIB-20) message was introduced to carry the scheduling information for one SC-MCCH per cell, that contains scheduling information for one SC-MTCH per each multicast service. When a new SC-PTM service becomes available in a cell, SC-MCCH is changed, therefore devices have to read SIB-20 to update the SC-MCCH. To inform devices about the changes in the SIB-20 network needs to broadcast SIB-1 messages (Option A in Fig.~\ref{fig:MtMSpro}). 
\begin{figure}[htbp]
	\centering
	\includegraphics[scale=0.45]{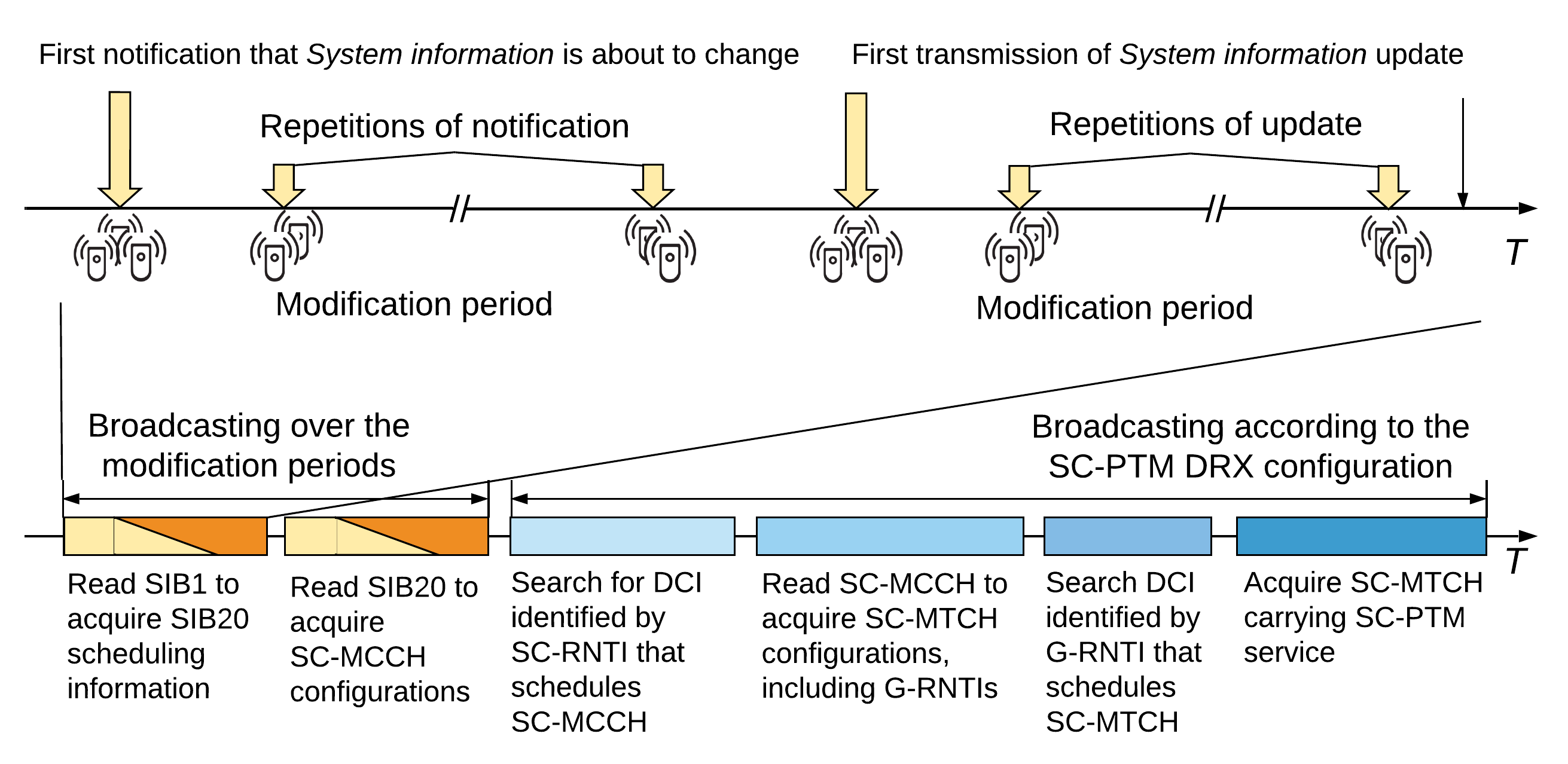}
	\caption{Delay of the standard SC-PTM transmission.} 
	\label{fig:MP}
\end{figure}

The transmission of one SIB message takes 64 frames or 640 ms~\cite{3GPP45820}. 
Notifications of SIB changes apply the concept of \textit{modification period}. It means that the system information content is not supposed to change within a modification period, and the same information can be repeated within a modification period. 
In the next modification period, the content is allowed to change.
Hence, during the first modification period, the BS informs devices that the information is about to change, but the updated information itself is transmitted only in the next modification period, as shown in Fig.~\ref{fig:MP}.

\begin{figure}[htbp]
	\centering
	\includegraphics[scale=0.5]{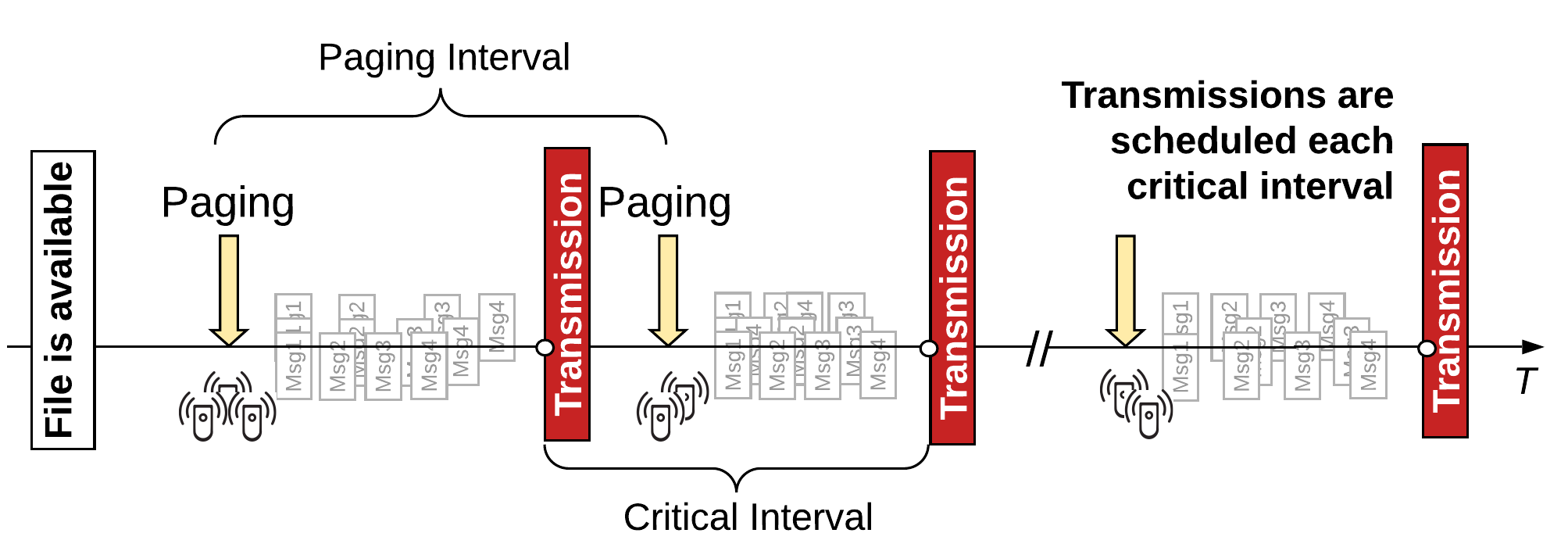}
	\caption{Paging and Multiple-subgroups Multicast Transmissions.}
	\label{fig:Prop}
\end{figure}

As we discussed in our previous work~\cite{BMSB}, the payload of critical IoT applications is relatively small and content must be delivered to devices with a minimal delay. The wait-for-all approach fails to fit such a requirement when the number of involved devices is high. 
We propose to send paging messages to small subgroups of devices and schedule multicast transmission in a short interval after paging as illustrated in Fig.~\ref{fig:Prop}. 

\begin{figure}[htb]	
    \centering
	\includegraphics[scale=0.6]{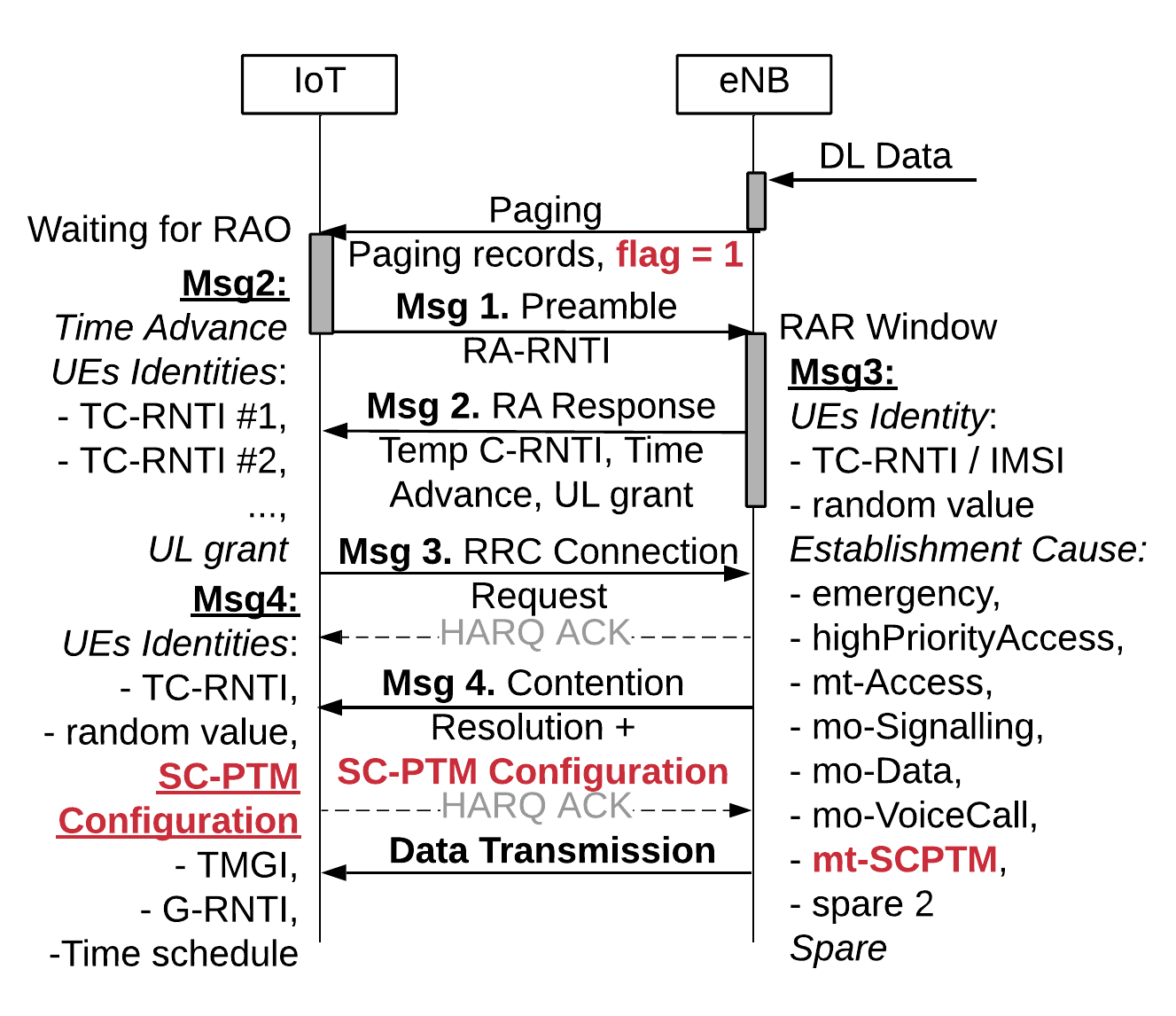}
	\caption{Enhanced RA procedure for the group-based critical communication.}
	\label{fig:RACH2}
\end{figure}

Upon receiving the list of relevant IoT devices, the network starts paging. All successfully paged devices have to initiate the RA procedure. The SC-PTM configuration information will be piggybacked on the Msg4 replacing the \textit{RRC Connection Setup/Resume} message. Fig.~\ref{fig:RACH2} illustrates the necessary modifications to the paging message and to the 3GPP compliant RA procedure to enable the proposed solution. A \textit{flag} in the paging message should be set to 1 to inform devices of the SC-PTM related paging. To emphasize that the SC-PTM configuration is requested, also Msg3 is extended to let device specify a new establishment cause in the corresponding spare field of Msg3 that we define as \textit{mt-Multicast}. When carrying SC-PTM configuration in Msg4, IoT devices benefit from the hybrid automatic repeat request (HARQ) mechanism that improve the reliability of the multicast service. However, the RA stage could be a bottleneck. Paging a large number of IoT devices may cause preamble retransmissions due to the limited opportunities for sending \textit{Msg2}, and may delay the RA completion. The less devices complete RA before the next scheduled SC-PTM transmission, the less devices join the multicast group. When the multicast subgroups are small, radio spectrum is not efficiently utilized and the total SC-PTM service delay increases.

We propose to page a relatively small number of IoT devices to ensure that all of them complete the RA stage before the SC-PTM transmission. Moreover, the next group of devices is paged only at the end of the RA stage of the previous group. The interval between two successive SC-PTM transmissions depends on the expected access delay and SC-PTM transmission delay. More details are given in the next section.

\section{System Model}
\label{sec:model}

\begin{figure}[tbh]
	\centering
	\includegraphics[scale=0.5]{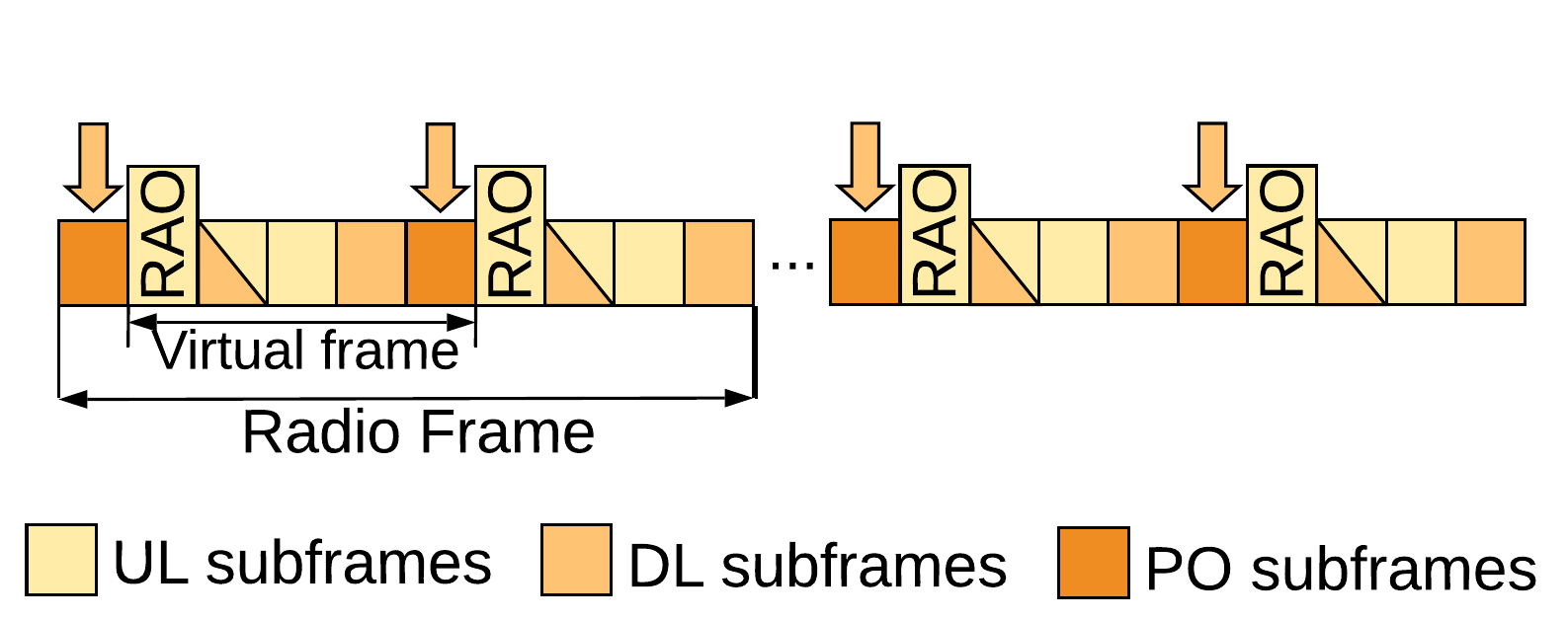}
	\caption{System time model.}
	\label{fig:VF}
\end{figure}

We consider a single-cell scenario with $N$ uniformly distributed devices.
Let us define a \textit{virtual frame (VF)} composed of $T_{VF}$ subframes as the time interval between two successive RAOs. 
The system time $T$ is slotted into $I = \lceil T/T_{VF}\rceil$ VFs, where $\mathcal{I} = \{1,\dots, I\}$ denotes VF indexes. We assume that each VF has one PO and one RAO, as illustrated in Fig.~\ref{fig:VF}.

Let $Q$ denote the number of paging subgroups, $\mathcal{Q} = \{1,\dots,Q\}$. If paging subgroup $q \in \mathcal{Q}$ has $n_{q}$ devices, then $n_1 + \ldots + n_Q = N$ and $n_q \leq N_{j}$, where $j \in \mathcal{J}$ denotes one of the paging schemes under consideration.

Let $\mathbf{P} = (\vec{P}_1,\ldots, \vec{P}_I)^T$ be the paging matrix composed of vectors $\vec{P}_i = (p_{i,q})_{i \in \mathcal{I}, q \in \mathcal{Q}}$, whose element $p_{i,q}$ denotes the number of devices in the paging subgroup $q$ at the VF $i$.
For a paging scheme $j \in \mathcal{J}$, we define $\mathcal{I}_j \subset \mathcal{I}$ as the subset of VF indexes in which paging messages should be sent. 
In particular, $\mathcal{J} = \{SP, GP, eGP, NeGP\}$. For the SP scheme, $\lceil N/N_{SP}\rceil$ VFs carry paging messages, where $N_{SP}=16$ and paging interval is equal to one VF, therefore  $\mathcal{I}_{SP} = (1,2,\ldots,\lceil N/N_{SP}\rceil)$. According to the GP scheme, all $N_{GP}=N$ devices can be reached by one paging message~\cite{GP}, so  $\mathcal{I}_{GP}$ consists of only one element. The eGP scheme claims that a new paging group ($N_{eGP}=36$) can be formed every $T_{eGP}=30$ ms, i.e., every $i_{eGP} = \lceil T_{eGP}/T_{VF}\rceil$ VFs, thus $\mathcal{I}_{eGP} = \{1, 1 + i_{eGP},\ldots, 1 + ([N/N_{eGP}] - 1)i_{eGP}\}$. 

In our proposed NeGP, we define $\mathcal{I}_{NeGP}$ by taking into account the RA and SC-PTM transmission delays. Specifically, $F$ VFs are needed to complete the 4-message handshake for the RA when $N_{NeGP}$ devices contend at the preamble transmission stage. Then, let $W$ denote the number of VFs required for the SC-PTM transmission. Thus, a new group of devices can be paged every paging interval $T_{NeGP} = (F + W)\cdot T_{VF} = i_{NeGP}\cdot T_{VF}$ ms, and $\mathcal{I}_{NeGP} = \{1, 1 + i_{NeGP},\ldots, 1 + ([N/N_{NeGP}] - 1)i_{NeGP}\}$. 
The optimal number of devices in a paging group is equal to the maximum number of devices that can be acknowledged in \textit{Msg2} during the RAR window, i.e. $N_{NeGP}=N_{RAR}$. By considering the RA control overhead of $\sigma = 30\%$ and the RAR message format~\cite{3GPP36331}, the maximum number of devices that can be acknowledged during the RAR window is computed as $N_{RAR} = [(1 - \sigma)D_0]\lceil T_{RAR}/T_{VF}\rceil$, where $D_0$ is the number of RBs available for the DL transmission in a VF, and $T_{RAR}$ the RAR window duration. For a given system configuration $D_0 = 12$ and $T_{RAR} = T_{VF}$, which yields $N_{NeGP} = 8$.

An IoT device that receives a paging message in VF $i$ initiates the RA at the same VF. If the first RA attempt fails, the device may take up to $R$ attempts, $\mathcal{R}=\{1,\ldots, R + 1\}$.
Let	vector $\vec{\alpha}_{i,r}$ denote the number of devices having the RA attempt $r$ in VF $i$, where $i \in \mathcal{I}$, $r \in \mathcal{R}$.

When devices make the first RA attempt, i.e. $r = 1$,
\begin{equation}
\label{eq:eq1}
\vec{\alpha}_{i,1} = \vec{P}_i, i \in \mathcal{I}. 
\end{equation}

The total number $\alpha_{i}$ of devices having Msg1 transmission in VF $i$ can be obtained as follows:
\begin{equation}
\label{eq:eq2}
\alpha_{i} = \left( \sum_{r = 1}^{R}{(\vec{\alpha}_{i,r})} \right) \cdot \mathbf{1}, i \in \mathcal{I},
\end{equation}
where $\mathbf{1} = ( 1,1,\dots,1 )^{T}$, $|\mathbf{1}| = Q$.

The random access to $C$ preambles by $\alpha_i$ devices is an instance of the occupancy problem. The probability to pick a preamble by a device from $C$ available preambles is equal to $1/C$. If $\alpha_i$ devices contend at VF $i$ the probability $q_i(c)$ of using exactly $c$ out of $C$ preambles at least by one device can be given as in~\cite{prob}:
\begin{equation}
\label{eq:eq3}
q_{i}(c) = \binom{C}{C - c} \sum_{j = 0}^{c}\left( - 1 \right)^{j}\binom{c}{j} \left( 1 - \frac{C - c + j}{C} \right)^{\alpha_{i}}.
\end{equation}

The expected number of used preambles $C_i$ in VF $i$, $i \in \mathcal{I}$, can be calculated as follows:
\begin{equation}
\label{eq:eq4}
C_{i} =  \Bigg[\displaystyle\sum_{c = 1}^{C_i^*} c q_{i}(c) \Bigg/ \displaystyle\sum_{c = 1}^{C_i^*} q_{i}(c) \Bigg]
\end{equation}
where $C_i^* = \min{(C,\alpha_i)}$. We normalize $\sum_{c = 1}^{C_i^*} c q_{i}(c)$ because the sum of probabilities $q_{i}(c)$ for $c = \{1,\dots, C_i^*\}$ does not hold $1$ when the number of contending devices $\alpha_{i}$ is less than $C$.
The probability $p_i$ of choosing a unique preamble in VF $i$ depends on the number of contending devices $\alpha_{i}$: 
\begin{equation}
\label{eq:eq5}
p_i = \left(1 - \frac{1}{C}\right)^{\alpha_i - 1}, i \in \mathcal{I}.
\end{equation}

Collided devices which have received the same UL grant in Msg2 collide again in Msg3 transmission and can repeat the RA attempt after the Contention Resolution Time (CRT) window expiration. We denote $M = \left\lceil T_{CRT}/T_{VF} \right\rceil$ as the CRT window $T_{CRT}$ in number of VFs.

The expected number of contending devices in VF $i$ is the total number of devices that make the first RA attempt after paging, devices that failed to receive Msg2, and devices that collided at step 3 of the RA procedure. Let $\vec{\alpha}^{*}_{i,r}$ denote the number of devices that successfully received Msg2 in VF $i$ after $r$ RA attempts. Vectors $\vec{\beta}_{i,r}$ and $\vec{\beta}_{i,r}^{*}$ stand for the number of devices scheduled for the Msg3 transmission in VF $i$ and for the number of devices that successfully sent Msg3 in VF $i$ after $r$ RA attempts, respectively. Finally, let $\vec{\gamma}_{i,r,m}$ denote the number of devices that receive Msg4 in VF $i$ after $m$ VFs of the contention resolution time and $r$ RA attempts, while $\vec{\gamma}_{i,r,m}^{*}$ stands for number of devices that successfully received Msg4 in VF $i$.

Devices that failed the RA attempt retry after the back-off window (BW) $T_{BW}$ or $j$ VFs, $j = \overline{1,B}$, where $B = \left\lceil T_{BW}/T_{VF} \right\rceil$. Let $\varphi_{j} = 1/B$ be the probability of randomly choosing the back-off time. 
The expected number of devices contending in VF $i$ yields:
\begin{align}
\label{eq:eq7}
&\vec{\alpha}_{i,r} = H[i - 1] \left( \vec{\gamma}_{i - 1,r - 1,M} - \vec{\gamma}_{i - 1,r - 1,M}^{*} \right) + H[i - k - M] p_{i - k - M} \cdot \vec{\beta}_{i - M,r - 1} + \nonumber\\
&+ \sum_{j = 1}^{B} H[i - j - 1] \left( \vec{\alpha}_{i - j - 1,r - 1} - \vec{\alpha}_{i - j - 1,r - 1}^{*} \right) \varphi_{j}, i \in \mathcal{I}, r \in \mathcal{R},j = \overline{1,B}
\end{align}
where $H[x]$ is a Heaviside function; it equals to 1 if $x > 0$ and takes 0 if $x\leq 0$.

The BS needs $T_{RA}$ ms to detect and decode transmitted preambles before sending Msg2. Thus, a device waits for  $k = \left\lceil ((A - 1)T_{VF} + T_{RA})/T_{VF}A \right\rceil$ VFs for the Msg2 reception. Let $N_{RAR}$ denote the system capacity for Msg2 transmissions in numbers of preambles that can be acknowledged by the BS. If devices contending in VF ($i - k$) used less than $N_{RAR}$ preambles, then all devices receive Msg2. Otherwise, only a portion of them receives Msg2, that is given as follows:
\begin{equation}
\label{eq:eq6}
\vec{\alpha}_{i,r}^{*} = \begin{cases}
\vec{\alpha}_{i - k,r}, & C_{i - k} \leq N_{RAR} \\
\left\lbrack \vec{\alpha}_{i - k,r} N_{RAR}/C_{i - k} \right\rbrack, & C_{i - k} > N_{RAR}. \\
\end{cases} 
\end{equation}

The expected number of devices to be scheduled for the Msg3 transmission in VF $i$ can be given as follows:
\begin{equation}
\label{eq:eq10}
\vec{\beta}_{i,r} = \vec{\alpha}_{i - 1,r}^{*} + \left(\vec{\beta}_{i - 1,r} - \vec{\beta}_{i - 1,r}^{*}\right),
\end{equation}
where $\left(\vec{\beta}_{i - 1,r} - \vec{\beta}_{i - 1,r}^{*}\right)$ counts for the devices that failed to send Msg3 in VF $i-1$ due to the lack of UL resources. 

Let $U_{0}$ be the total number of UL resources available in VF $i$. Since the PRACH occupies a fixed number $U_{P}$ of RBs in the UL, the number of available UL resources in VF $i$ for Msg3 transmission equals to $U_{i} = U_{0} - U_{P}$.
The expected number of devices scheduled for the Msg3 transmission in VF $i$ can be given as follows: 
\begin{equation}
\label{eq:eq11}
\vec{\beta}_{i,r}^{*} = \begin{cases}
\vec{\beta}_{i,r}, & \vec{\beta}_{i,r}\mathbf{u}^{T} \leq U_{i} \\
\left\lbrack \vec{\beta}_{i,r}U_{i}/\vec{\beta}_{i,r} \mathbf{u}^{T} \right\rbrack, & otherwise, \\
\end{cases}  
\end{equation}
where $\mathbf{u}^{T}$, $|\mathbf{u}| = Q$, denotes the average number of RBs required for the Msg3 transmission.

The expected number of devices to be scheduled for the Msg4 transmission in VF $i$ is either the number of devices that successfully sent Msg3 in the previous VF or the number of devices that failed to receive Msg4 in the previous VF due to the lack of the DL resources:
\begin{equation}
\label{eq:eq12}
\vec{\gamma}_{i,r,m} = \begin{cases}
\vec{\beta}_{i - 1,r}^{*}, & m = 1 \\
\vec{\gamma}_{i - 1,r,m - 1} - \vec{\gamma}_{i - 1,r,m - 1}^{*}, & otherwise. \\
\end{cases} 
\end{equation}
where $i \in \mathcal{I}$, $r \in \mathcal{R}\setminus \{R + 1\}$.

Let $D_{0}$ and $D_{RAR}$ be the total number of DL resources available in VF $i$ and the average number of resources required for the Msg2 transmission, respectively. The number of DL resources $D_i$ after the Msg2 transmission can be calculated as:
\begin{equation}
\label{eq:eq8}
D_{i} = \begin{cases}
D_{0} - D_{RAR}, &\left( \displaystyle\sum_{r = 1}^{R}\vec{\beta}_{i,r} \right) \mathbf{1}^{T} > 0 \\
D_{0}, & otherwise.
\end{cases}
\end{equation}
Therefore, the expected number of devices that successfully sent Msg4 in VF $i$ yields:
\begin{equation}
\label{eq:eq13}
\vec{\gamma}_{i,r,m}^{*} = \begin{cases}
\vec{\gamma}_{i,r,m}, & \vec{\gamma}_{i,r,m}\mathbf{d}^{T} \leq D_{i}\  \\
\left\lbrack \vec{\gamma}_{i,r,m} D_{i}/\vec{\gamma}_{i,r,m}\mathbf{d}^{T} \right\rbrack, & otherwise, \\
\end{cases} 
\end{equation}
where $\textbf{d}^{T}$ denotes the average number of DL resources required for the Msg4 transmission, $|\mathbf{d}| = Q$.

After receiving Msg4 in VF $i$, devices can receive SC-PTM transmission scheduled in one of the next VFs. We assume that up to $S$ multicast transmissions can be scheduled within $I$ VFs, $\mathcal{S} = \{1,\ldots, S\}$. Let $i_s$ be the first VF of the SC-PTM transmission $s$. Then, the expected number $\vec{\delta_s}$ of devices ready for the SC-PTM transmission $s$ yields:
\begin{equation}
\label{eq:eq20}
\vec{\delta}_s = \displaystyle\sum_{k = i_{s - 1}}^{i_{s} - 1}{\displaystyle\sum_{r = 1}^{R}{\sum_{m = 1}^{M}{\vec{\gamma}_{k,r,m}^{*}}}}, s \in \mathcal{S}.
\end{equation}

Let $z$ define the critical interval between two successive SC-PTM transmissions. The first transmission should be scheduled with an offset to ensure that all devices of the first paging subgroup receive Msg4, while all next multicast transmissions are scheduled in $z$ VFs.

Let $\Theta$ be the multicast payload in terms of resources needed for the SC-PTM transmission. The residual number of resources $\theta_{l_s}$ required to complete transmission $s$ after the first $l_s - 1$ VFs is given as follows:
\begin{equation}
\label{eq:eq19}
\theta_{l_s} = \begin{cases}
\Theta, & l_s = 0 \\
\theta_{l_s-1} - D_{i^{*}_s + l_s}, & \theta_{l_s-1} > D_{i^{*}_s + l_s} \\
0, & otherwise.\\
\end{cases}
\end{equation}
Let $l_s^*$ stands for the last VF of the SC-PTM transmission $s$ such that $\theta_{l^{*}_{s}}=0$, i.e. denotes the duration of the SC-PTM transmission $s$. The expected number of devices $\vec{\delta}^{*}_{s}$ that successfully receive the multicast service after $l_{s}^{*}$ VFs equals to $\vec{\delta}_{s}$. We now can calculate the metrics of interests.

\textit{Access success probability} $P_{A}$ is a ratio of the number of devices that completed the RA stage to the overall number of devices reached through paging
\begin{equation}
\label{eq:eq21}
P_{A} = 1 - \left( \displaystyle\sum_{i = 1}^{I}\vec{\alpha}_{i,R + 1} \right)\mathbf{1}^{T}/\left( \displaystyle\sum_{i = 1}^{I}\vec{\alpha}_{i,1} \right)\mathbf{1}^{T}.
\end{equation}

\textit{Average access delay} $D_{A}$ corresponds to the time to complete the RA:
\begin{equation}
\label{eq:eq22}
D_{A} = \frac{1}{Q}\displaystyle\sum_{q = 1}^{Q}{\left(i^*_q - i_q \right)T_{VF}},
\end{equation}
where $i_{q}$ stands for the VF at which group $q$ receives paging and 
$i^*_q$ is given as follows
\begin{equation}
\label{eq:eq23}
i^*_q = \left[	 
	\left( \displaystyle\sum_{i = 1}^{I}{i} 
	\displaystyle\sum_{r = 1}^{R}
	\displaystyle\sum_{m = 1}^{M} \vec{\gamma}_{i,r,m}^{*} 
	\right) \mathbf{e}_q^T \Big/ 
	\left( 
	\displaystyle\sum_{i = 1}^{I} \vec{\alpha}_{i,1} 
	\right) \mathbf{e}_q^T
\right].
\end{equation}

\textit{Average idle delay} $D_{Idle}$ is the time that elapses from the end of the RA stage until the beginning of the multicast transmission, therefore 
\begin{equation}
\label{eq:eq25}
D_{Idle} = \frac{1}{Q}\displaystyle\sum_{q = 1}^{Q}{
	\left(i^{**}_q - i^*_q\right)}T_{VF}.
\end{equation}
where $i_{q}^{**}$ is given as follows
\begin{equation}
\label{eq:eq26}
i^{**}_q = \left\lbrack
\displaystyle\sum_{s\in \mathcal{S}}{i^*_s}\left(\vec{\delta}_{s}\mathbf{e}_q^T\right)
\Big/ \left( \displaystyle\sum_{s \in \mathcal{S}}\vec{\delta}_{s} \right)\mathbf{e}^T_q 
\right\rbrack - 1
\end{equation}
because not all devices of the same paging subgroup will be members of the same multicast subgroup for the SC-PTM reception.

\textit{Average total delay} $D_{Total}$ includes the average access delay $D_{A}$, average idle delay $D_{Idle}$, and average SC-PTM transmission delay $D_{TX}$: 
\begin{equation}
\label{eq:eq28}
D_{Total} = D_{A} + D_{Idle} + D_{TX},
\end{equation}
where the average SC-PTM transmission delay can be computed as 
\begin{equation}
\label{eq:eq30}
D_{TX} = \frac{1}{S} \displaystyle\sum_{s = 1}^{S}{l^*_s} \cdot T_{VF}.
\end{equation}

\textit{Total service delay} $D_{Service}$ is the total time to wake up all relevant devices and deliver the content of interest. 
Having $i_{S^*}$ and $l_{S^*}$ of the very last multicast transmission $S^*$, we compute the metric as follows
\begin{equation}
\label{eq:eq33}
D_{Service} = (i_{S^*} + l_{S^*}) T_{VF}.
\end{equation}

\textit{Average access energy consumption} $E_A$ can be given as an arithmetic mean of the average energy consumption per paging subgroup $E_{A_q}$:
\begin{equation}
\label{eq:eq37}
E_{A} = \frac{1}{Q}\displaystyle\sum_{q = 1}^{Q}{E_{A_q}}.
\end{equation}

Let $t_1$, $t_2$, $t_3$ and $t_4$ be the average transmission delay of Msg1, Msg2, Msg3 and Msg4. The device energy consumption in transmission mode equals to $e_{TX}$ mW, in reception mode - $e_{RX}$ mW, devices in idle mode consume $e_{Idle}$ mW on average. In the access stage, devices of subgroup $q$ consume:
\begin{align}
\label{eq:eq34}
E_{A_q} = &(e_{TX}t_{1} + e_{RX}t_{2}) r_q^2 + (e_{TX}t_{1} + e_{RX}t_{2} + e_{TX}t_{3}) (r_q^3 + 1) +  \nonumber\\
&+ e_{Idle} T_{BW} r_q^2 + e_{RX}t_4,
\end{align}
where $r_q^2$ and $r_q^3$ denote the average number of retransmission attempts due to failure after Msg2 and Msg3 transmission, respectively. The average number of RA attempts due to Msg2 or Msg3 failure is computed as the weighted mean:
\begin{equation}
\label{eq:eq35}
r_q^2 = \frac{ \left(
	\displaystyle\sum_{r = 1}^{R} r 
	\displaystyle\sum_{i = 1}^{I} \left( \vec{\alpha}_{i,r} - \vec{\alpha}^{*}_{i,r} \right) 
	\right) \mathbf{e}_q^T}
{\left( 
	\displaystyle\sum_{r = 1}^{R}
	\displaystyle\sum_{i = 1}^{I}\left( \vec{\alpha}_{i,r} - \vec{\alpha}^{*}_{i,r} \right)
	\right) \mathbf{e}_q^T}.
\end{equation}
\begin{equation}
\label{eq:eq36}
r_q^3 = \frac{ \left(
	\displaystyle\sum_{r = 1}^{R} r 
	\displaystyle\sum_{i = 1}^{I} \vec{\alpha}^s_{i,r} \left(1 - p_i \right) 
	\right)\mathbf{e}_q^T}{ \left(
	\displaystyle\sum_{r = 1}^{R}
	\displaystyle\sum_{i = 1}^{I} \vec{\alpha}^s_{i,r} \left(1 - p_i \right) 
	\right) \mathbf{e}_q^T}.
\end{equation}

\textit{Average device energy consumption} is the total energy consumed during the access, idle and SC-PTM transmission stages by a device on average: 
\begin{equation}
\label{eq:eq39}
E_{Total} = \left(E_A + e_{Idle} D_{Idle} + e_{TX} D_{TX}\right).
\end{equation}

\textit{Resource utilization} $R_{UL}$ and $R_{DL}$ is the ratio between the number of occupied resources and the total number of available resources in $I$ VFs in the UL and DL, respectively:
\begin{equation}
\label{eq:eq40}
R_{UL} = 1 - \frac{\sum_{i = 1}^{I}{U_i}}{I U_0},
\end{equation}
\begin{equation}
\label{eq:eq41}
R_{DL} = 1 - \frac{\sum_{i = 1}^{I}{D_i}}{I D_0}.
\end{equation}

\section{Selected numerical results}
\label{sec:results}

\begin{table}[htb!]
\footnotesize
	\centering
	\caption{Reference system model parameters}
	\label{tab:model_setup}
	\begin{tabular}{p{1.4cm}p{8.3cm}p{2.4cm}}
		\hline
		Notation & Definition & Value \\ \hline \hline
		$C$ & Number of available preambles & 54 \\
		$R$ & Maximum number of preamble retransmissions & 10 \\
		$N_{j}$ & Paging group size, $j=\{SP, GP, eGP, NeGP\}$ & $\{16, N, 36, 8\}$ \\
		$T_{j}$ & Paging interval, $j = \{SP, GP, eGP, NeGP\}$ & $\{5, 0, 30, 25\}$ ms\\
		$A$ & Number of RA subframes in a radio frame & 2 \\
		$d$ & Interval between two consecutive POs & 5 ms \\
		$z$ & Critical interval & 25 ms \\
		$T_{VF}$ & Virtual frame duration & 5 ms \\
		$T_{RA}$ & Delay for the preamble detection and decoding & 5 ms \\
		$T_{RAR}$ & RAR window & 5 ms \\
		$T_{BW}$ & Back-off window & 20 ms \\
		$T_{CRT}$ & Contention resolution time & 48 ms\\
		$N_{RAR}$ & Number of devices that may receive RAR within $T_{RAR}$ & 8 \\
		$U_0$ & Amount of resources available for the uplink transmission in each VF & 12 RBs \\
		$U_{P}$ & Amount of resources occupied by PRACH in the UL & 12 RBs \\
		$D_0$ & Amount of resources available for the downlink transmission in each VF & 12 RBs \\
		$D_{RAR}$ & Amount of resources required for the RAR message transmission in DL VF $i$ & 6 RBs \\
		$\mathbf{u}$ & Vector of the average number of resources for Msg3 transmission & (1,\ldots,1) RBs \\
		$\mathbf{d}$ & Vector of the average number of resources for Msg4 transmission & (1,\ldots,1) RBs \\
		$\Theta$ & Multicast traffic payload & $\{3,12,32\}$ RBs \\
		$e_{Tx}$ & Average device power consumption in the transmit mode & 500 mW \\
		$e_{Rx}$ & Average device power consumption in the receive mode & 80 mW \\
		$e_{Idle}$ & Average device power consumption in the idle mode & 3 mW \\
		\hline
	\end{tabular}
\end{table}

\begin{table}[ht]
\footnotesize
	\centering
	\caption{Simulation parameters}
	\label{tab:sim_setup}
	\begin{tabular}{p{4.6cm}p{8cm}}
		\hline
		Parameter & Value  \\
		\hline \hline
		Cell radius & 500 m \\
		Carrier configuration & 1.4 MHz carrier bandwidth at 800 MHz \\
		PHY numerology & TDD frame type 1, TTI 1 ms \\
		RA capacity & 2 RAOs per radio frame \\
		Resource allocation & PDSCH, PDCCH: 1 -- 6 PRBs \\
		& PUSCH, PUCCH: 1 -- 6 PRB, \\
		& PRACH: 6 PRBs \\ 
		Device power class & 23 dBm \\
		BS transmit power & 46 dBm \\
		Power consumption & 500 mW (TX), 80 mW (RX), 3mW (Idle)\\
		Traffic payload & $\{392, 1608, 4584\}$ bits \\
		\hline
	\end{tabular}
\end{table}

We compare our paging solution, named \textit{New enhanced Group Paging (NeGP)}, over three reference paging strategies, namely \textit{Standard Paging (SP)}~\cite{SP} (i.e. legacy 3GPP solution), \textit{Group Paging (GP)}~\cite{GP}, and \textit{enhanced Group Paging} (eGP)\cite{Condoluci}. 

We consider a symmetric radio frame configuration (with the same number of UL and DL subframes) with $A=2$ RAOs, as shown in Fig.~\ref{fig:VF}. The mentioned paging strategies have different number of devices per paging subgroup and different paging intervals. For the reader's convenience, we give definitions of the system model parameters and their corresponding values in Table~\ref{tab:model_setup}.
The analytic results have been validated by simulations in MATLAB. Simulation parameters are set according to~\cite{3GPP45820} and~\cite{3GPP36213}, for radio interface, and to~\cite{Power}, for device energy consumption, as reported in Table \ref{tab:sim_setup}. Data packets arriving in a burst of a given size are transmitted over a set of continuous subframes.

In the following figures, analytical results are shown as solid lines with markers, and simulation results only as markers; an almost perfect match is observed. Results are plotted for a cluster of up to 500 devices camping on a single LTE-M narrowband. As explained in~\cite{SP}, the device arrival rate of 40.3 access attempts per second with a target outage probability below 1\% corresponds to the LTE-M trafﬁc capacity per narrowband equaled to $0,36 \cdot 10^6$ devices/km$^2$, the higher capacity of $10^6$ devices/km$^2$ can be achieved if three or more narrowbands are conﬁgured in a cell. Our NeGP paging solution allows 320 device arrivals per second with outage probability less than 1\%, which ensures more than $10^6$ device/km$^2$ of supported connection density.
\begin{figure}[tbh]
	\centering
	\subfigure[Average access delay]{\includegraphics[scale=0.35]{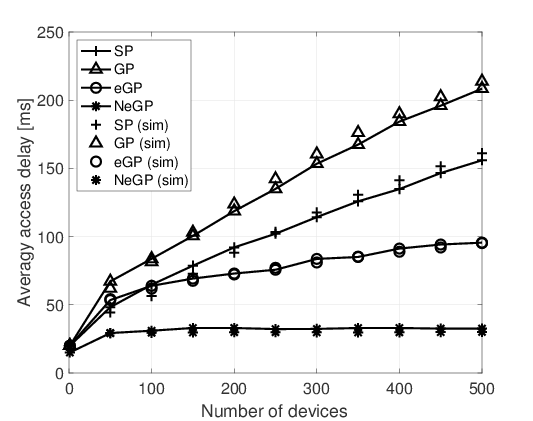}}
	\subfigure[Average access energy consumption]{\includegraphics[scale=0.34]{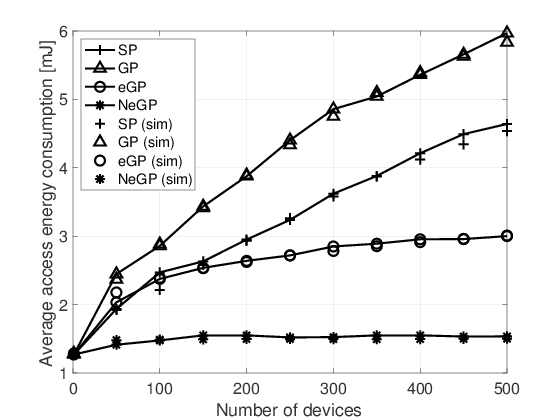}}
	\caption{Average access delay and energy consumption.}
	\label{fig:AD}
\end{figure}

Fig.~\ref{fig:AD} shows the average access delay (a) and average device energy consumption (b) for different paging strategies. 
The GP scheme introduces a significant delay and energy usage at the RA stage with respect to other schemes due to the high number of contending devices. For the SP and GP schemes both metrics grow almost linearly when the number of devices increases due to the preamble collisions and lack of radio resources. On the contrary, both metrics tend to saturate in the cases of the eGP and NeGP schemes. The eGP solution exploits the code-expanded preamble transmission technique that decreases collision rate and, consequently, the number of preamble retransmission attempts~\cite{Condoluci}. However, our NeGP solution shows more than 50\% reduction of both the average access delay and the average device energy consumption compared to the eGP scheme. 
The reason behind such performance gain is that the size of the paging groups and paging intervals in NeGP are well customised in such a way that devices complete the RA without any additional delay caused by preamble collisions or shortage of the radio resources.
\begin{figure}[tbh]
	\centering
	\subfigure[Average idle delay]{\includegraphics[scale=0.34]{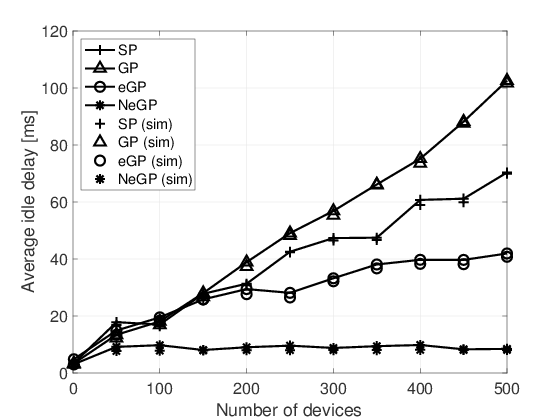}}
	\subfigure[Average device energy consumption]{\includegraphics[scale=0.34]{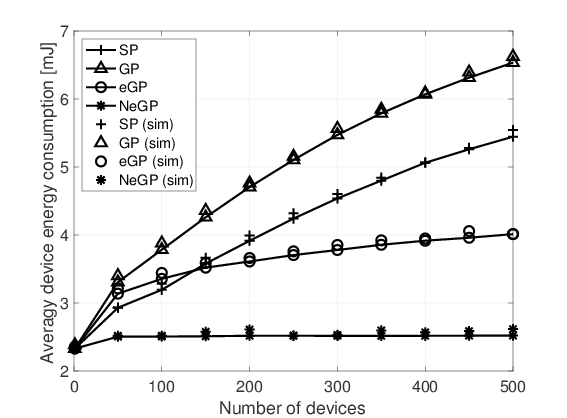}}
	\caption{Average idle delay and average device energy consumption.}
	\label{fig:WD}
\end{figure}

Devices that complete the RA procedure remain in idle mode while waiting for the SC-PTM transmission but keep listening to the DL since the last transmission until the end of the Inactivity timer defined by the DRX. If the timer expires before the SC-PTM transmission, devices switch off their receiving antenna and become unavailable until the next PO.
Fig.~\ref{fig:WD}(a) shows the average idle delay, i.e. the time to wait for the SC-PTM transmission after the reception of SC-PTM configuration parameters. The idle delay of the GP scheme grows fast under increasing number of devices. In the case of SP and eGP, the metric increases mainly due to the short paging interval or high number of devices per group. To ensure that all paged devices receive the multicast transmission, the Inactivity timer should be higher than the idle delay. Fig.~\ref{fig:WD}(b) illustrates the average device energy consumption under the assumption that the Inactivity Timer is set according to the experienced idle delay. The metric constantly grows under GP, SP and eGP strategies but it is almost constant for the NeGP scheme. This is an important result for battery-powered IoT devices. 

\begin{figure}[htbp]
	\centering
	\subfigure[Small payload]
	{\includegraphics[scale=0.34]{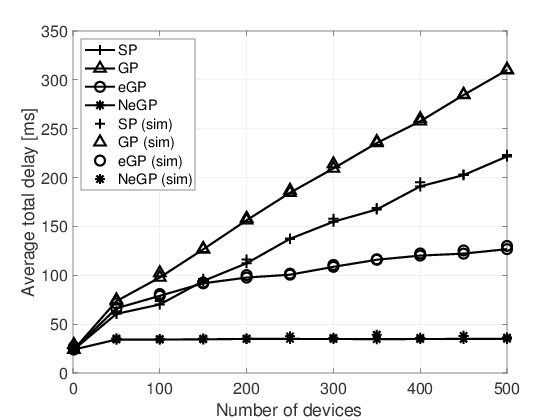}}
	\subfigure[Large payload]
	{\includegraphics[scale=0.34]{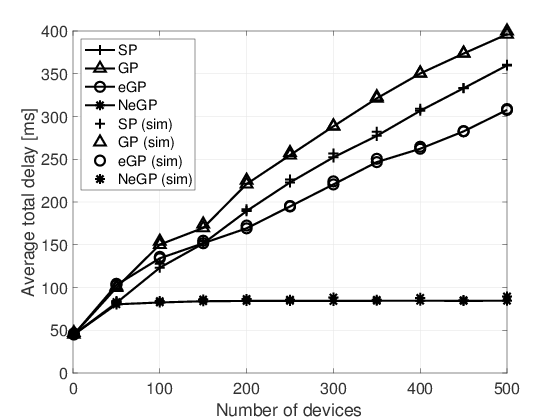}}
	\caption{Average total delay in case of: (a) small payload, 
	and (b) large payload.}
	\label{fig:TD}
\end{figure}

Fig.~\ref{fig:TD} shows the average total delay for the variable SC-PTM payload. In particular, the size is set to 392, 
and 4584 bits. For simplicity, we refer to these values as small (a) 
and large (b) payload, respectively. The total delay includes access delay, idle delay and the time to transmit SC-PTM payload. 
The system performance is sensitive to the payload size because long multicast transmissions may overlap with the RA stage. Our NeGP paging and SC-PTM transmission design has been designed in order to avoid such an overlapping. 
As shown in Fig.~\ref{fig:TD}, the increase of SC-PTM payload does not lead to the significant performance degradation in the case of NeGP and results only in an additional deterministic delay.

\begin{figure}[htbp]
	\centering
	{\includegraphics[scale=0.34]{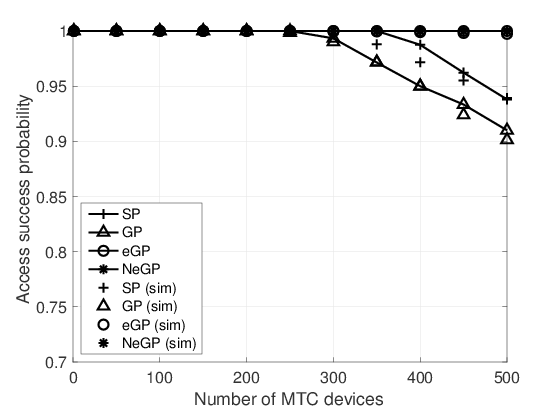}}
	\caption{Access success probability.}
	\label{fig:ASP}
\end{figure}

The access success probability is shown in Fig.~\ref{fig:ASP}(a). This metric also can be used as the \textit{service probability} if necessary assumptions on the Inactivity Timer are made, as previously discussed. The failures are not only caused by preamble collisions but also by retransmissions after Msg2 and Msg3 failures. When the number of devices in the SP and GP schemes is increased not all devices can successfully complete the RA. For a cluster of 500 devices, from 5\% to 10\% of devices fail the RA in the case of SP and GP strategies. Very few devices lose the SC-PTM transmission if the eGP scheme is applied, while the NeGP guarantees the successful completion of the RA procedure by all devices. 

\begin{figure}[htbh]
	\centering
	\subfigure[small]{\includegraphics[scale=0.33]{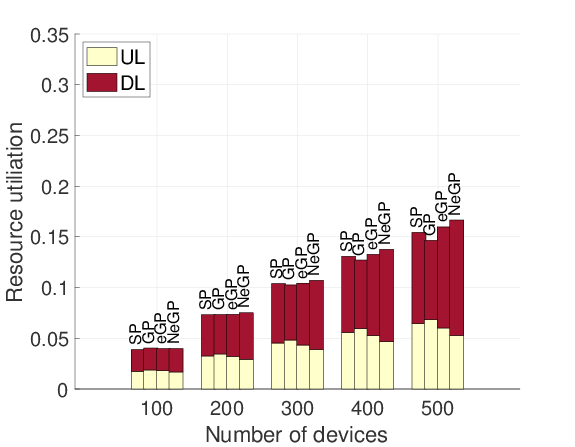}}
	\subfigure[large]{\includegraphics[scale=0.33]{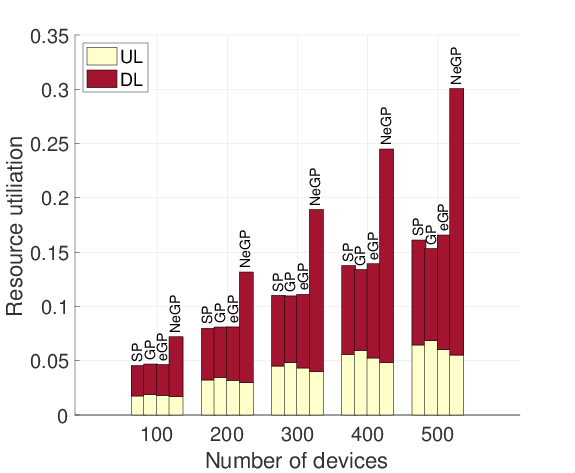}}
	\caption{UL (yellow) and DL (red) resources utilization in the case of (a) small 
	and (b) large payload.}
	\label{fig:RU}
\end{figure}

We compare the performance of our proposal with reference schemes in terms of radio resource consumption in the UL and DL for different payloads as reported in Fig.~\ref{fig:RU}. 
Regarding the UL utilization, the NeGP scheme requires less resources than SP, GP and eGP solutions, because it does not incur retransmissions of the RA messages. On the contrary, GP requires more UL resources than any other paging strategy due to the higher collision rate. Having more UL resources available is advantageous for the system that can support other background traffic (e.g., from other IoT devices).
The DL resource utilization depends on the number of multicast transmissions required to service all relevant devices. As expected, the NeGP solution requires more DL resources because it induces more SC-PTM transmissions. The difference in required DL resources becomes more evident when the payload size is larger and more devices wait for the multicast service.

\section{Conclusions}
\label{sec:conclusions}

In this paper, we investigated a wide set of performance metrics to evaluate the proposed multicast framework for the delivery of initially unplanned critical multicast traffic towards bandwidth- and power-limited cIoT devices. We proposed to schedule identical SC-PTM transmissions over an finely tuned interval to improve the service probability and reduce device energy consumption. 
We extensively compared our solution over similar reference schemes, both analytically and via simulations. 
We highlighted that paging significantly impacts the performance of critical SC-PTM communication when the arrival of multicast traffic can not be predicted. 
The optimal configuration of paging and SC-PTM scheduling guarantees 100\% of the service delivery and stable device total delay irrespective of the number of receivers but at the expense of a long service delay. However, a short device total delay is more preferable than a short service delay in critical applications.

\end{document}